\documentclass[a4paper]{JHEP3}

\usepackage{graphicx}
\usepackage{amsmath}
\usepackage{cite}

\def\eqref#1{(\ref{#1})}
\def\text{\rm }

\def\samurai{{{\sc samurai}}}
\def\cuttools{{{\tt CutTools}}}
\def\golem{{{\tt Golem\;95}}}
\newcommand{\bite}{\begin{itemize}}
\newcommand{\eite}{\end{itemize}}

\newcommand{\beq}{\begin{equation}}
\newcommand{\eeq}{\end{equation}}
\newcommand{\bqa}{\begin{eqnarray}}
\newcommand{\eqa}{\end{eqnarray}}

\def\db#1{\bar D_{#1}}

\def\slh#1{\rlap / {#1}}

\def\spa#1.#2{\langle#1\,#2\rangle}
\def\spb#1.#2{[#1\,#2]}

\def\spab#1.#2.#3{\langle\mskip-1mu{#1}
                  | #2 | {#3}]}

\def\spba#1.#2.#3{[\mskip-1mu{#1}
                  | #2 | {#3}\rangle}

\def\spbb#1.#2.#3.#4{[\mskip-1mu{#1}
                     | {#2} \ {#3} | {#4}]}

\def\spaa#1.#2.#3.#4{\langle\mskip-1mu{#1}
                     | {#2} \ {#3} | {#4}\rangle}

\newcommand{\bea}{\begin{eqnarray}}

\newcommand{\eea}{\end{eqnarray}}

\newcommand{\bean}{\begin{eqnarray*}}

\newcommand{\eean}{\end{eqnarray*}}
           
\newcommand{\nn}{\nonumber \\}

\title{Tensorial Reconstruction at the Integrand Level}

\author{G. Heinrich \\
IPPP, Department of Physics, University of Durham, Durham DH1 3LE, UK;\\
Institute for Theoretical Physics, University of Z\"urich, Winterthurerstr. 190, \\
8057 Z\"urich, Switzerland\\
E-mail: \email{gudrun.heinrich@durham.ac.uk}}

\author{G. Ossola \\ 
Physics Department, New York City College of Technology,\\ 
                      City University Of New York, 
           300 Jay Street, Brooklyn NY 11201, USA.\\
E-mail: \email{GOssola@citytech.cuny.edu}
}

\author{T. Reiter \\
Nikhef, Science Park 105, 1098XG Amsterdam, The Netherlands. \\
E-mail: \email{thomasr@nikhef.nl}
}

\author{F. Tramontano \\
Theory Group, Physics Department, CERN,
           CH-1211 Geneva 23, Switzerland.\\
E-mail: \email{francesco.tramontano@cern.ch}
}

\preprint{CERN-PH-TH/2010-175 \\ Nikhef~2010-020\\IPPP/10/65\\
DCPT/10/130\\ZU-TH 10/10}

\abstract{We present a new approach to the reduction of one-loop amplitudes obtained
by reconstructing the tensorial expression of the scattering amplitudes. 
The reconstruction is performed at the integrand level by means of a sampling in the integration momentum.
There are several interesting applications of this novel method within existing techniques for the reduction of
one-loop multi-leg amplitudes: to deal with numerically unstable points, such as in the vicinity of a vanishing Gram determinant; 
to allow for a sampling of the numerator function based on real values of the integration momentum; to optimize the numerical reduction in the case of long expressions for the numerator functions.
}

\begin{document}

\section{Introduction}

In the last few years we observed enormous progress in the computation of 
one-loop virtual corrections for processes involving many particles.

Commonly, all these calculations rely on the decomposition of the tensor
integrals in terms of scalar master integrals which are known 
analytically, and can be computed by
means of public libraries, like LoopTools~\cite{Hahn:2010zi}, QCDloop~\cite{Ellis:2007qk,vanOldenborgh:1989wn}, \golem~\cite{Binoth:2008uq} or  OneLOop~\cite{vanHameren:2009dr,vanHameren:2010cp}. 
The task of reducing tensor integrals to scalar integrals can be achieved by 
following a fully analytic procedure, according to the well established 
Passarino-Veltman reduction~\cite{thv, pv}, which proved its effectiveness for reactions involving a small number of particles. Improved tensor reduction methods led to the development of tools~\cite{Binoth:2005ff,Denner:2005nn,Binoth:2008uq,Denner:2010tr} which are able to deal in an efficient and numerically stable way with processes of high complexity such as 
the EW corrections to $e^+ e^- \to 4 {\rm f}$~\cite{Denner:2005fg}
and the QCD corrections to
$pp\to t\bar t b\bar b$~\cite{Bredenstein:2009aj,Bredenstein:2010rs,Bredenstein:2010je},
or  $ q \bar q \to b\bar b b\bar b$~\cite{Binoth:2009rv}.

In recent years, inspired by unitarity arguments~\cite{Bern:1994zx, Bern:1994cg}, alternative approaches have emerged,
aiming at the direct determination of the coefficients of the master integrals. 
A novel powerful framework for one-loop calculations was developed by merging 
the idea of exploiting the kinematic cuts of the scattering amplitudes in four dimensions \cite{Britto:2004nc,Bern:2007dw}
with a systematic analysis of the structure of their integrands~\cite{delAguila:2004nf}, leading to a framework by now
known as OPP method~\cite{Ossola:2006us,Ossola:2007bb}. This approach has been further extended to the generalized
$d$-dimensional unitarity-based method~\cite{Ellis:2007br,Giele:2008ve} for the 
reconstruction of scattering amplitudes in dimensional regularization~\cite{Ellis:2008ir}. 

Different versions of the new reduction techniques have been implemented in various codes, such as \cuttools~\cite{cuttools}
and \samurai~\cite{samurai}, which are publicly available, BlackHat~\cite{Berger:2008sj}, and Rocket~\cite{Giele:2008bc}.
Among several results obtained with these new approaches~\cite{Binoth:2008kt,Actis:2009uq,Melnikov:2010iu,Biswas:2010sa},
the state of the art is represented by the numerical 
calculation of extremely challenging $2\to 4$ processes, like $pp\to W+3$~jet
production~\cite{Berger:2009zg,Berger:2009ep,Ellis:2009zw,KeithEllis:2009bu},
$pp\to Z+3$~jet production~\cite{Berger:2010vm}, $pp\to t\bar t b\bar b$~\cite{Bevilacqua:2009zn},
$pp\to t\bar t jj$~\cite{Bevilacqua:2010ve}, and $pp\to W^+ W^+ jj$~\cite{Melia:2010bm}.

Controlling the quality of the numerical reconstruction is crucial for the new approaches to one-loop
calculations~\cite{cuttools,samurai,Pittau:2010tk}: with respect to the tensorial reduction, the new
approaches require a particular attention to the issues of numerical efficiency and the control over numerical instabilities,
which are some of the strong features of purely algebraic techniques. 
In addition to standard tests, commonly used within all the reduction methods 
to monitor the quality of the final results
(such as the checks on the coefficients of the UV and IR poles 
or the comparison between results 
using different levels of numerical precision), 
the OPP/$d$-dimensional unitarity framework provides internal-consistency checks on the quality of the reconstructed coefficients. 
The phase space points for which the reconstruction fails the tests
are in general (re)processed using routines that provide higher numerical precision
which demand longer computing time.
Multi-precision procedures are implemented either in dedicated routines or
by doubling the original algorithm within a copy running in higher floating 
point precision.

In this paper we present a new simple strategy for the numerical evaluation 
of one-loop amplitudes, based on the reconstruction of the tensorial representation 
of their integrands.
Accordingly, any integrand can be expressed in a basis of  tensors
formed by products of the integration momentum, each multiplied by 
a tensorial coefficient.
The tensorial decomposition is achieved by sampling the 
integrand for {\it real} values of the components of the loop momentum.
Finally, the reconstructed coefficients can be contracted numerically with the corresponding tensor integrals, evaluated by means of dedicated libraries such as \golem~\cite{Binoth:2008uq}.

This technique shows its virtues in at least three cases.

First, it is suitable to treat unstable configurations in the proximity of
vanishing Gram determinants.
The reduction of tensor integrals to express them 
in terms of scalar ones is responsible
for the appearance of spurious, potentially singular coefficients, 
which might spoil the numerical accuracy
in phase space regions where the Gram determinants tend to zero.
Using a tensor basis, rather than a scalar one, this problem is 
bypassed completely.
Although the evaluation of the tensor integrals demands more time than
the computation of the scalar integrals, it might still be competitive when compared to
a complete reduction in higher floating point precision.
Moreover, a ``rescue-system" based on the tensorial reconstruction of the integrand 
should be performed only for a small fraction of points
when integrating over the full phase~space.

Second, the tensorial reconstruction combined with the sampling 
by {\it real} momenta yields the automatic generation of the integrand 
available directly from tree-level generators like
MadGraph~\cite{Alwall:2007st} and HELAC~\cite{Kanaki:2000ey, Cafarella:2007pc},
which  provide tree amplitudes for real values of external momenta.
This possibility can be an alternative to the unitarity-based generation 
of the integrand, which, by construction, requires tree-level amplitudes 
evaluated at complex (yet on-shell) external momenta.

A third possibility is represented by the use of
the reconstructed tensor integrand as ``preprocessed'' integrand for the
reduction algorithm.
Namely, the tensorial reconstruction
of the integrand, by definition, amounts to 
disentangling the part of the integrand depending on the loop momentum
from the one depending  only on the kinematic variables.
The dependence on the kinematics is carried by the tensorial coefficients, 
which have to be evaluated only once per phase-space point.
Therefore, given any one-loop integrand, one can reconstruct its tensorial structure
at a given phase-space point, and use the reconstructed expression as input for
the reduction procedure at the integrand level.
The numerical evaluation of the  preprocessed integrand can be more effective 
than the evaluation of the original integrand, because the kinematic information
stored in the tensorial coefficients is constant during the reduction algorithm, 
as it evolves 
simply through the variation of the loop-momentum used for the sampling.
Within the framework of the preprocessed numerator, the tensorial coefficients
play the role of an algebraic alternative to a caching system. 

The paper is organized as follows. The reconstruction algorithm is discussed in Section~\ref{sec:tensred}.
Section~\ref{sec:appl} describes the various applications of the method, with particular emphasis on the ``rescue-system" option. 
In this Section we also describe an example of implementation by means of existing codes such as \samurai{} and \golem{}. Finally in Section~\ref{sec:end} we present our conclusions.

\section{Tensorial Reconstruction Algorithm}
\label{sec:tensred}

\subsection{Algorithm}

Within the dimensional regularization scheme, any one-loop $n$-point amplitude 
can be written as
\bea
&& {\cal A}_n = \int d^d {\bar q} \frac{{\cal N}({\bar q}, \epsilon)}{\db{0}\db{1}\cdots \db{n-1}} \ , \nn
&& \db{i} = ({\bar q} + p_i)^2-m_i^2 = 
            (q + p_i)^2-m_i^2-\mu^2 .
\label{def:An}
\eea
We use a bar to denote objects living in $d=~4-2\epsilon$  
dimensions, following the prescription
$\slh{{\bar q}} = \slh{q} + \slh{\mu}$ with ${\bar q}^2= q^2 - \mu^2$. 
Further, we use the notation $f({\bar q})$ as short-hand notation for $f(q,\mu^2)$.

Our task is to rewrite the numerator function ${\cal N}({\bar q})$ as a linear combination of tensors of
increasing rank, up to the maximum power of the integration momentum appearing in the numerator.

We first focus on the part of the numerator ${\cal N}({q})$ that depends only on the 4-dimensional
part of the loop momentum $q$. Let's assume that the numerator has at most $R$ powers in the integration momentum $q$. 
We aim at building numerically the tensorial representation which reproduces ${\cal N}({q})$ before integration for each value of $q$. In the most general case, we have an expression of the form
\begin{equation} \label{4dim}
\mathcal{N}(q)=
	\sum_{r=0}^{R}	C_{\mu_1\ldots\mu_r} q_{\mu_1} \ldots q_{\mu_r} \, ,
\end{equation}
where for each $r$ the set of coefficients $C_{\mu_1 \ldots \mu_r}$ forms a
contravariant tensor (we keep lower indices for convenience) and the
contraction is performed with the Euclidean metric $(1, 1, 1, 1)$.
For $r=0$, we indicate the constant term as $C_0$.
We observe that the reconstruction of Eq.~(\ref{4dim}), namely the calculation
of all coefficients $C_{\mu_1\ldots\mu_r}$,
is independent of the number of denominators $\db{i}$  appearing in the original amplitude.
It will however depend on the specific phase space point or helicity configuration that we want to process, 
in the same way as the numerator functions of other numerical methods.

In order to determine all the coefficients, we can simply evaluate both sides of Eq.~(\ref{4dim}) for an arbitrary set of 
values of the integration momentum. Those values can be chosen as real four-momenta, thus allowing the treatment of numerators  depending on a real integration momentum.

It is useful to rewrite ${\cal N}({q})$ by separating the tensorial components.
Each of the terms in Eq.~\eqref{4dim} can be written
as a multivariate polynomial in the components of~$q$,
where $q_4$ denotes the energy component
\begin{equation}
C_{\mu_1\ldots\mu_r}q_{\mu_1}\cdots q_{\mu_r}=
\sum_{(i_1, i_2, i_3, i_4)\vdash r}
\hat{C}^{(r)}_{i_1 \, i_2\, i_3\, i_4}\cdot
(q_1)^{i_1}(q_2)^{i_2}(q_3)^{i_3}(q_4)^{i_4}.
\label{eq:CvsChat}
\end{equation}
Here, the notation $\vdash$ indicates that the indices $i_j$ have to
form an integer partition of~$r$.

We can now compute the coefficients $\hat{C}^{(r)}$;
the conversion from $\hat{C}\mapsto C$ is easy since each component
of $C_{\mu_1\ldots\mu_r}$ contributes to one
particular $\hat{C}^{(r)}_{i_1\ldots i_4}$ where the tuple
$(\mu_1,\ldots,\mu_r)$ contains exactly $i_j$ occurrences
of the number~$j$. Conversely, when contracting the
tensor~$C_{\mu_1\ldots\mu_r}$
with a tensor integral of rank~$r$, one has to take into account that
$\hat{C}^{(r)}_{i_1\ldots i_4}$ already sums over
\begin{displaymath}
\frac{\left(\sum_{j=1}^4 i_j\right)!}{\prod_{j=1}^4 (i_j)!}
\end{displaymath}
symmetric components of~$C$.
For both $C$ and $\hat{C}^{(r)}$ the
number of independent components in four dimensions~is
\begin{equation}
n_r=\left({4+r-1}\atop r\right)
\end{equation}
and therefore $n_r=\{ 1, 4, 10, 20, 35, 56, 82 \}$ for $r=0, 1, \ldots 6$ respectively.

To reconstruct the tensorial coefficients of Eq.(\ref{4dim})
we will then consider the numerator for four-dimensional loop momentum $q$,
${\cal N}(q)$ $={\cal N}(x,y,z,w)$,
as a multivariate polynomial of degree $R$
in the four variables $x,y,z$ and $w$, being the components
of $q$, {\it i.e.} $q_\mu = (x,y,z,w)$.
We will determine all the coefficients by sampling $q$,
(hence its four components) in a bottom-up approach, 
starting from the simplest choice of $q$, namely the null-vector,
and proceeding with sets of momenta $q$ where one variable at a time is 
different from zero.

\subsubsection{Level-0}

The coefficient $C_0$ can be immediately determined 
from the identity:
\begin{equation}
\mathcal{N}(0,0,0,0) \equiv {\cal N}^{(0)}= C_0
\end{equation}
where $q = (0, 0, 0, 0)$. 
This trivially completes the calculation of the constant term.

\subsubsection{Level-1}

We can subtract ${\cal N}^{(0)}$ from ${\cal N}({q})$,
and define 
\begin{equation}
{\cal N}^{(1)}({q}) \equiv \mathcal{N}(q) - {\cal N}^{(0)} \ .
\end{equation}

In this case we may consider four polynomial systems, generated
respectively by choosing $q$ with only one non-vanishing component,
whose solution leads to the determination of $l_1=4R$ coefficients. \\
Using $q = (x, 0, 0, 0)$, we extract only the $R$ coefficients that 
are proportional to the first component of $q$. 
Therefore our system reduces to
\begin{equation} \label{sub1x} 
{\cal N}^{(1)}(x,0,0,0) \equiv 
x\, C_{1} + x^2\, C_{11} + \ldots + x^R\, C_{\underbrace{11 \ldots 1}_{R {\rm \ times}}} \, ,
\end{equation}
and can be easily solved by choosing $R$ different values for $x$. \\
We repeat the same operation for $q = (0,y,0,0)$,
\begin{equation} \label{sub1y} 
{\cal N}^{(1)}(0,y,0,0) \equiv 
y\, C_{2} + y^2\, C_{22} + \ldots + y^R\, C_{\underbrace{22 \ldots 2}_{R {\rm \ times}}} \, ,
\end{equation}
and proceed analogously with
the remaining two choices of $q$ with only one component different from zero.

\subsubsection{Level-2}

After completing Level-0 and Level-1, we know $4R+1$
coefficients, and we can again subtract the corresponding terms 
from $\mathcal{N}(q)$, to obtain
\begin{equation}
{\cal N}^{(2)}({q}) \equiv
{\cal N}^{(1)}({q}) 
- \sum_{j=1}^4 C_j \ q_j 
- \sum_{j=1}^4 C_{jj} \ q_j^2
- \ldots -\sum_{j=1}^4 C_{\underbrace{jj \ldots j}_{R {\rm \ times}}} \ q_j^R \, .
\end{equation}
We proceed with the calculation of the $l_2=3 R (R-1)$ coefficients multiplying two 
non-vanishing components of $q$. 
They will be found from six systems, each containing $R(R-1)/2$ coefficients. \\
We begin with $q= (x, y, 0, 0)$, and 
with this choice of $q$,  ${\cal N}^{(2)}({q})$ reduces to:
\begin{equation}
{\cal N}^{(2)}(x,y,0,0) \equiv
 x\, y\, C_{12} + x^2\, y\, C_{112} + x\, y^2\, C_{122} + \ldots 
+ \sum_{a,b: (a+b=R)} x^a\, y^b\, \hat{C}^{(R)}_{ab00} \ ,
\end{equation}
where $a$ and $b$ are different from zero.
To extract the coefficients, we solve straightforwardly a system generated 
by sampling $R(R-1)/2$ pairs $(x,y)$. \\
Then we choose $q=(x,0,z,0)$, and get the generating polynomial for the second
system,
\begin{equation}
{\cal N}^{(2)}(x,0,z,0) \equiv
 x\, z\, C_{13} + x^2\, z\, C_{113} + x\, z^2\, C_{133} + \ldots 
+ \sum_{a,b: (a+b=R)} x^a\, z^b\, \hat{C}^{(R)}_{a0b0} \ ,
\end{equation}
which yields again $R(R-1)/2$ coefficients.
We repeat the same sampling algorithm for the remaining four cases.

\subsubsection{Level-3}

The first step of Level-3 is the subtraction of the
known coefficients from ${\cal N}(q)$, defining the polynomial
\begin{equation}
{\cal N}^{(3)}(q) \equiv 
{\cal N}^{(2)}(q)
- \sum_{r=2}^R \ \sum_{a,b: (a+b)=r} \ \sum_{k<j=1}^4 
C_{\underbrace{j \ldots j}_{a {\rm \ times}}\underbrace{k\ldots k}_{b {\rm \ times}}} \ q_j^a q_k^b 
\end{equation}
This time, we proceed with the calculation of the $l_3=2 R (R-1) (R-2)/3$ coefficients by multiplying three non-vanishing components of $q$. 
They will be found from four systems, each containing $R(R-1)(R-2)/6$ coefficients. \\
We begin with $q= (x, y, z, 0)$, so that 
${\cal N}^{(3)}({q})$ reduces to:
\begin{equation}
{\cal N}^{(3)}(x,y,z,0) \equiv
 xyz \, C_{123} + x^2y z \, C_{1123} + \ldots 
+ \sum_{a,b,c:(a+b+c=R)} x^a y^b z^c\,\hat{C}^{(R)}_{abc0} \ ,
\end{equation}
where $a,b$ and $c$ are different from zero.
By sampling $R(R-1)(R-2)/6$ triplets $(x,y,z)$, we generate a system for
the extraction of the coefficients.
Analogous to level-2, we continue with the other three polynomials
generated when ${\cal N}^{(3)}({q})$ is evaluated respectively for
$q=(x,y,0,w)$, $q=(x,0,z,w)$, and $q=(0,y,z,w)$.
The solutions of the associated systems complete  the
reconstruction of all coefficients multiplying three non-vanishing components of $q$.

\subsubsection{Level-4}

The final step is represented by the
determination of the $l_4=R(R-1)(R-2)(R-3)/24$ coefficients multiplying
four non-vanishing components of $q$.
They  form the polynomial defined as
\begin{equation}
{\cal N}^{(4)}(q) \equiv 
{\cal N}^{(3)}(q)
- \sum_{r=3}^R \ \sum_{a,b,c: (a+b+c)=r} \ \sum_{m<k<j=1}^4 
C_{\underbrace{j \ldots j}_{a {\rm \ times}}
   \underbrace{k\ldots k}_{b {\rm \ times}}
   \underbrace{m\ldots m}_{c {\rm \ times}}
} \ q_j^a \, q_k^b \, q_m^c \ .
\end{equation}
This polynomial is evaluated at $q=(x,y,z,w)$, 
\begin{equation}
{\cal N}^{(4)}(x,y,z,w) \equiv
 xyzw \, C_{1234} + \ldots 
+ \sum_{a,b,c,d:(a+b+c+d=R)} x^a y^b z^c w^d\,\hat{C}^{(R)}_{abcd} \ ,
\end{equation}
($a,b,c$ and $d$ are different from zero) and sampled $R(R-1)(R-2)(R-3)/24$
times, to generate the system for the extraction on the last set of unknown
coefficients.

At the end of Level-4, the tensorial reconstruction is complete:
the original numerator ${\cal N}({q})$ is rewritten as a combination
of tensors formed by products of loop momenta $q$.
The tensorial coefficients multiplying each tensor store
the information depending on the kinematics.
We will denote the reconstructed numerator by 
${\langle N({q}) \rangle}$. 
For example,
in the case of a numerator ${\cal N}({q})$ of rank six, the 
reconstruction of ${\langle N({q}) \rangle}$ requires the calculation
of 210 coefficients, obtained by evaluating ${\cal N}({q})$ 210 times. 
More details on the combinatorics are  given in Appendix~A.
We would like to point out that by construction the proposed algorithm 
does not introduce any spurious sources of instabilities.\\
As already mentioned, ${\langle N({q}) \rangle}$ can be employed
in several ways. \\
First,  it can be used as {\it rescue system}
to treat unstable configurations in the proximity of
vanishing Gram determinants, because the use of 
a tensor basis, rather than a scalar one,
avoids the appearance of spurious potentially singular coefficients.\\
Second, as ${\langle N({q}) \rangle}$ can be constructed from products
of trees evaluated at {\it real} momenta only, it 
matches well with the automatic generation of the integrand 
directly from available tree-level generators like
MadGraph~\cite{Alwall:2007st} and HELAC~\cite{Kanaki:2000ey, Cafarella:2007pc},
which  provide tree amplitudes for real values of external momenta.
Third, 
the reconstructed tensor integrand ${\langle N({q}) \rangle}$ 
can be used as an  integrand which is more suitable for 
further processing, 
as will be explained in detail in section \ref{sec:numsampling}.

\subsection{Example of the reconstruction for rank two}

If the numerator has at most two powers of $q$, Eq.~(\ref{4dim}) reads
\begin{equation} \label{4dim:rank2}
\mathcal{N}(q)=
	C_0 + C_{\mu}q_{\mu} + C_{\mu\nu} q_{\mu} q_{\nu} \ .
\end{equation}
We have a total of 15 independent coefficients to determine,
because $q_{\mu} q_{\nu}$ selects the symmetric part of $C_{\mu \nu}$.
According to the algorithm outlined in the previous section,
to reconstruct the unknown coefficient we have to solve the following
systems.
\begin{itemize}

\item Level-0. \\
We start from ${\cal N}(q)$ and evaluate it at the null-vector,
\begin{equation}
{\cal N}(0,0,0,0)={\cal N}^{(0)} = C_0 \ ,
\end{equation}
obtaining trivially the constant $C_0$.

\item Level-1. \\
Subtracting the known term, we define
\begin{equation}
{\cal N}^{(1)}({q}) \equiv \mathcal{N}(q) - {\cal N}^{(0)} \ ,
\end{equation}
and evaluate it for the following four choices of $q$,
\begin{eqnarray}
{\cal N}^{(1)}(x,0,0,0) &=& x \ C_{1} + x^2 \ C_{11} \ , \\
{\cal N}^{(1)}(0,y,0,0) &=& y \ C_{2} + y^2 \ C_{22} \ , \\
{\cal N}^{(1)}(0,0,z,0) &=& z \ C_{3} + z^2 \ C_{33} \ , \\
{\cal N}^{(1)}(0,0,0,w) &=& w \ C_{4} + w^2 \ C_{44} \ .
\end{eqnarray}
Sampling each polynomial for two different {\it real} values
of the variable, and solving the corresponding systems,
yields the determination of the $l_1=4R=8$ unknown coefficients $C_i$ and $C_{ii}$
with $i=1,\ldots,4$.

\item Level-2. \\
We again subtract the determined coefficients,  define
\begin{equation}
{\cal N}^{(2)}({q}) \equiv
{\cal N}^{(1)}({q}) 
- \sum_{j=1}^4 C_j \ q_j 
- \sum_{j=1}^4 C_{jj} \ q_j^2
\end{equation}
and evaluate it for the following six choices of $q$,
\begin{eqnarray}
{\cal N}^{(2)}(x,y,0,0) &=& xy \ C_{12} \ , \\
{\cal N}^{(2)}(x,0,z,0) &=& xz \ C_{13} \ , \\
{\cal N}^{(2)}(x,0,0,w) &=& xw \ C_{14} \ , \\
{\cal N}^{(2)}(0,y,z,0) &=& yz \ C_{23} \ , \\
{\cal N}^{(2)}(0,y,0,w) &=& yw \ C_{24} \ ,  \\
{\cal N}^{(2)}(0,0,z,w) &=& zw \ C_{34} \ . 
\end{eqnarray}
The extraction of the $l_2=3R(R-1) =6$ coefficients is trivial 
in this case as well,
because each monomial has to be sampled only once.

\end{itemize}

At the end of Level-2, we have found the numerical values
of the 15 coefficients appearing in Eq.(\ref{4dim:rank2}).

\subsection{Reconstruction of the $\mu^2$-dependence}
\label{sec:tensred:d}

To account for the $\mu^2$-dependence of the numerator ${\cal N}({\bar q})$,
where $\mu^2$ is the radial integration variable in the 
$(d-4)$-dimensional subspace, 
the tensorial decomposition of the numerator requires more terms
than in the 4-dimensional case  previously discussed.
The extended decomposition reads
\begin{equation} \label{ddim} 
{\cal N}({\bar q}) =  
{\langle N({q}) \rangle} 
+ G^{(1)} \mu^2 
+ G^{(2)} \mu^4 + G^{(3)}_\alpha q^\alpha \mu^2
+ G^{(4)}_{\alpha \beta} q^\alpha q^\beta \mu^2 \, .
\end{equation}
This form can be derived from the fact that rational terms 
only come from the combination of
$(d-4)$-dimensional terms with UV divergent integrals,
and the listed numerators are the only ones leading to UV divergent integrals
in addition to ${\langle N({q}) \rangle}$ (in a renormalisable~gauge).
Note that the $\mu^2$ terms can be inferred from the general relation 
\bqa
\int \frac{d^{d}q}{i \pi^{\frac{d}{2}}}
 \,(\mu^2)^\alpha \,f(q^\mu,q^2) 
 & = &
 \frac{\Gamma(\alpha+\frac{d}{2}-2)}{\Gamma(\frac{d}{2}-2)}
 \int \frac{d^{d+2\alpha}q}{i \pi^{\frac{d}{2}+\alpha}} f(q^\mu,q^2)\;.
 \label{ktilde}
\eqa
In calculating the terms proportional to both powers of $q$ and $\mu^2$, we should consider that not all terms will contribute
to the final result. We will therefore limit our procedure to non-vanishing terms plus additional vanishing terms which are required to compute other pieces. 
For example, by power counting, we can exclude the presence of the term $G^{(2)} \mu^4$ from
bubble and triangle diagrams. On the other hand, we cannot neglect the term $G^{(1)} \mu^2$ for box diagrams, even if we know that it will give a vanishing contribution: since we are operating at the integrand level, its presence is necessary in order to compute
other non-vanishing terms in Eq.~\eqref{ddim:4p}, starting from $G^{(2)}$.
Moreover, only the diagonal terms of $G^{(4)}$ in Eq.~\eqref{ddim:4p} are needed, since the non-vanishing part of the resulting integral is proportional to $g^{\alpha\beta}$.
Therefore, 
for $n$ denominators, the numerator function ${\cal N}_n({\bar q})$ of Eq.~\eqref{ddim} will reduce to:
\begin{eqnarray} 
\label{ddim:2p} 
{\cal N}_2({\bar q}) &=&  {\langle N({q}) \rangle} + G^{(1)} \mu^2 \ , \\
\label{ddim:3p} 
{\cal N}_3({\bar q}) &=&  {\langle N({q}) \rangle} + G^{(1)} \mu^2 
+ G^{(3)}_\alpha q^\alpha \mu^2 \ , \\
\label{ddim:4p} 
{\cal N}_4({\bar q}) &=&  {\langle N({q}) \rangle} + G^{(1)} \mu^2 
+ G^{(2)} \mu^4 + G^{(3)}_\alpha q^\alpha \mu^2
+ G^{(4)}_{\alpha \alpha} q^\alpha q^\alpha \mu^2 \ , 
\end{eqnarray}
respectively.

The presence of $\mu^2$ amounts to dealing with multivariate polynomials
with one more variable than in the four-dimensional case.
We can reconstruct all the coefficients $G^{(i)}$ by applying a technique
similar to the one employed in the previous Section, therefore,
we will not repeat the discussion on the adopted sampling.

The integrals in $d$ dimensions necessary for the evaluation of these terms have been given in several papers, see e.g.~\cite{Bern:1995db,vanHameren:2005ed,Binoth:2006hk,Draggiotis:2009yb}; we list them here for completeness:
\bqa 
\int d^d \bar{q}
\frac{\mu^2}{\db{i}\db{j}}             &=& - \frac{i \pi^2}{2}
\left[m_i^2+m_j^2-\frac{(p_i-p_j)^2}{3} \right]   +
\cal{O}(\epsilon)\,, \nn 
\int d^d \bar{q}
\frac{\mu^2}{\db{i}\db{j}\db{k}}       &=& - \frac{i \pi^2}{2} +
\cal{O}(\epsilon)\,,\nn
\int d^d \bar{q}
\frac{\mu^4}{\db{i}\db{j}\db{k} \db{l}} &=& - \frac{i \pi^2}{6} +
\cal{O}(\epsilon)\,, \nn
\int d^d \bar{q}
\frac{\mu^2\, q^{\mu}}{\db{i}\db{j}\db{k}} &=& \frac{i \pi^2}{6} (p_i+p_j+p_k) +
\cal{O}(\epsilon)\,,\nn
\int d^d \bar{q}
\frac{\mu^2\, q^{\mu}\, q^{\nu} }{\db{i}\db{j}\db{k} \db{l}} &=& - \frac{i \pi^2}{12} \, g^{\mu \nu} +
\cal{O}(\epsilon)\,.
\eqa 

Their multiplication by the coefficients $G^{(i)}$ completes the calculation of the rational part  corresponding to the evaluation of $R_2$ in Refs.~\cite{Ossola:2008xq,Draggiotis:2009yb,Garzelli:2009is}. 
After this step is done, we can set $\mu^2 = 0$ in all expressions and retain only the four-dimensional numerator in the remainder of the calculation.

\section{Examples of Applications}
\label{sec:appl}

The tensorial reconstruction, paired with an efficient program for the evaluation of tensor integrals, can 
represent a simple and efficient way of computing one-loop
virtual corrections. This method is particularly efficient for processes  involving integrands of maximum rank equal to four: in this case the number of coefficients to compute is small and routines for the evaluation of tensor integrals very efficient. Important processes such as $pp \to b\bar b b\bar b$ or $pp \to t\bar t b\bar b$ belong to this category.

In this Section however we explore different uses of the tensorial decomposition, in particular the advantages that this method can bring when combined with other  advanced  approaches for the reduction of one-loop amplitudes, such as OPP/$d$-dimensional unitarity.

\subsection{Dealing with unstable points}

One of the major challenges for the new reduction methods, in particular when compared to algebraic reduction techniques, is to deal efficiently and automatically with phase space points which are numerically unstable.
This is typically in the proximity of a vanishing Gram determinant. 

All the points that do not pass the reconstruction/stability test within any chosen reduction algorithm can be reprocessed by
using the technique presented here. The tensorial reconstruction avoids the reduction to scalar integrals and thus the emergence of Gram determinants; on the other hand, the technique requires the evaluation of tensor integrals, which is in general more time consuming. 

In order to approach continuously a kinematic configuration that is numerically unstable,
we consider a 4-point rank 4 diagram made of a fermion loop with two massless and 
two massive vector particles attached to it.
We approach the phase space configuration of a vanishing Gram determinant
by taking the limit $Q\rightarrow 0$ within the kinematics
\begin{align*}
p_{1,2} &=(E,0,0,\pm E) \quad \quad \quad \quad \quad \quad \quad \,\, p^2_{1,2}= 0 \\
p_{3,4} &=(E,0,\pm Q \sin\theta ,\pm Q \cos\theta) \quad \quad  p^2_{3,4} = m^2
\end{align*}
where $E=\sqrt{m^2+Q^2}$ changes with $Q$, while $\theta$ and $m^2$ are kept constant
(in the following plots we set $\theta=\frac{35}{180}\pi$ and $m^2=7$).
The Gram determinant is given by $\det G=32\,E^4Q^2\sin^2\theta$, while $\det S$, the modified Cayley determinant, goes to a constant as $Q\rightarrow 0$.
\begin{figure}[hbt]
\begin{center}
\includegraphics[width=0.3\textwidth]{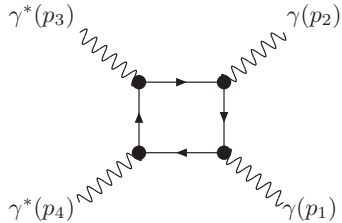}
\end{center}
\caption{For the example in this section we use  the
above diagram in QED with  two massless and two off-shell photons 
attached to  a massless fermion loop.}\label{fig:gram:photons}
\end{figure}

The calculation is performed with {\samurai}, using a standard $d$-dimensional
integrand-level reduction; the results are compared to the improved tensorial
technique where we employ \golem{} for the evaluation of the tensor integrals.

We present the case of a four-point function because this is typically 
the case in which the Gram determinant issue shows up.
We would like to note that for kinematics 
far from a vanishing Gram determinant, the new method based on
tensorial reconstruction and \samurai{} have similar performances;
however,  for diagrams with more than four legs, 
the performance of \samurai{} 
with ``standard" reduction at the integral level is naturally better in these 
``safe" phase space regions.

\begin{figure}[htbp]
\begin{center}
\includegraphics[width=0.8\textwidth]{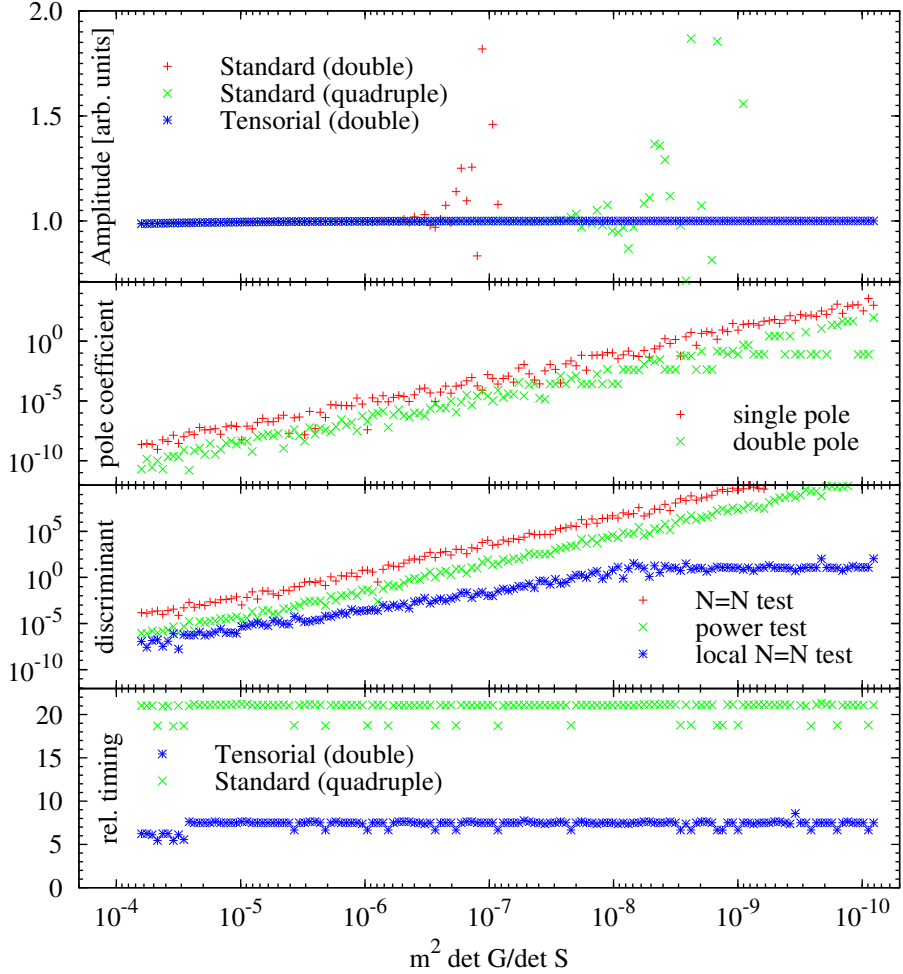}
\end{center}
\caption{Top panel: Comparison between standard reduction 
at the integrand level (Standard)
and tensorial reconstruction combined with an evaluation of the tensor
integrals (Tensorial). The standard method starts deviating from the correct
result at $\det G/\det S\approx10^{-7}$, the standard method with quadruple precision
(see the text) starts deviating at $\det G/\det S\approx10^{-8}$, 
while the tensorial method remains stable over the whole range.
Middle panels: the behaviour of tests that trigger the detection of
instabilities.
Bottom panel: timing of tensorial reduction in double precision 
versus standard reduction in quadruple precision,
normalized by the timing of standard reduction in double precision.
}\label{fig:gram:value}
\end{figure}

As shown in the top panel of Figure~\ref{fig:gram:value}, at a certain value
of~$\det G/\det S$, the 
standard reduction technique starts deviating from the stable result obtained by the
improved tensor reconstruction, exhibiting the sensitivity to the vanishing
Gram determinant.

The price to pay for the improved stability  is  an
increased run time.
The bottom panel of Figure~\ref{fig:gram:value} shows the timing
evaluated for the tensorial reconstruction method in double precision, compared to an
implementation in quadruple precision of the standard reduction at the integral level,
each normalized to the standard reduction in double precision.

To make a comparison with quadruple or multiple precision we follow
a strategy similar to the one proposed in \cite{Pittau:2010tk}: we generate the
kinematics in double precision, the scalar integral evaluation is performed
in double precision as well. Then we upgrade formally
the double precision kinematics to quadruple and go through the
algorithm evaluating the numerator with multiple precision.

This can certainly give some benefit because
near a configuration of vanishing Gram determinant the situation is 
such that numerator and denominator
vanish, and with quadruple precision one can succeed to get a cancellation
for small values of the Gram determinant.
Such an improvement is evident from Figure~\ref{fig:gram:value}
(top panel); nevertheless,
within this construction we do not find further improvements
increasing the precision of the algorithm from quadruple to multiple.
Hence, this implementation of quadruple precision only delays
the Gram determinant problem, thus reducing the
affected part of the phase space rather than curing the problem at its root.
One might argue that a little improvement here would mean a big improvement
in terms of statistics when extensive Monte Carlo runs are performed.
On the other hand, in our implementation,
examining the timing indicates that quadruple precision is more time
consuming then tensorial reconstruction,
at least for the example considered here.

The performance of the tensorial reconstruction relies on the underlying implementation
of the tensor integrals: for this purpose we employed the \golem{} library.
The timing depends on the value of $\left\vert\frac{\det G}{\det S}\right\vert$,
as expected given the internal structure of the \golem{} library.
When $\left\vert\frac{\det G}{\det S}\right\vert$
is above a certain threshold a fast, analytic reduction
is used. For points where the ratio is below the threshold, the
library switches to a method using the numerical quadrature of
one-dimensional integral representations of the tensor integral
form factors, thus avoiding the introduction of inverses of Gram
determinants, but increasing the run time by up to a factor seven
as compared to the standard reduction at integrand level,
 for this particular example.

Considering that the number of points where
the standard approach fails in a cross-section calculation
typically amounts to a few per mil, the additional cost is relatively
small. Moreover, within the set of diagrams contributing to a specific process, only a subset will
suffer from the instabilities and requires the additional evaluation time. 

We have so far only discussed how the proposed method can cure
numerical instabilities related to small Gram determinants, and will
now elucidate strategies for an a priori detection of points requiring
the application of this reduction technique.
One would like to keep the fraction of points which are classified
as unstable as low as possible to avoid switching unnecessarily to a
slower method. 
In the remainder of this section we will study the performance in detecting
such points for various reconstruction tests that have been proposed in the
literature.

\paragraph{Cancellation of the poles.}
The cancellation of the infrared and ultraviolet poles is often used
as a first indicator to identify points suffering from numerical instabilities.
The panel below the top one of Figure~\ref{fig:gram:value} shows that there is indeed
a correlation between the size of the Gram determinant and the deviation
of the poles from the expected result,
the latter being obtained from the tensorial method
which gives a stable answer.
However, one also observes a sizable spread in the values, which 
hampers the choice of a good threshold to discriminate unstable from stable
points.

\paragraph{The ``$N=N$'' local test.}
The local test only examines the reconstructed polynomials corresponding to specific cuts. In the example we have plotted at each point the bubble cut
which shows the maximum discriminant, the latter being defined as
 the relative difference between the reconstructed and the original polynomial.
We find a steep slope in the region
where the standard reduction method fails, however, the curve flattens out
at smaller values of the Gram determinant.

\paragraph{The ``$N=N$'' global test.}
A stronger correlation can be found by looking at the global ``$N=N$'' reconstruction
test. The idea behind this test is to compare the original
numerator function to the reconstructed one for an arbitrary value of
the integration momentum. 
The global test shows a clear correlation to the
Gram determinant. By comparing the top plot of Figure~\ref{fig:gram:value} with the benchmark from the 
``$N=N$'' global test of Figure~\ref{fig:gram:value}, we can select an appropriate threshold:
in this particular example, a threshold of ${\cal }O(1)$ in the discriminant is sufficient 
for the classification of points  as ``unstable".

\paragraph{The \emph{power test}.}
The so called power test was first presented in~\cite{samurai}. It uses
the fact that in the reconstruction of the numerator one typically allows
for a higher power of the integration momentum in the reconstructed
function than one would expect in the original numerator function.
One can therefore find certain combinations of coefficients which
should sum to zero if the reconstruction was successful.
Moreover, with respect to other reconstruction tests, the power test has the
advantage of being totally independent from the choice of the integration momentum.
Similar to the case of the global $N=N$ test we find a strong correlation
between the goodness of these zeroes and the size of the Gram determinant
and therefore find another trustworthy criterion for the discrimination
of numerically unstable points in the presence of degenerate kinematics.

\subsection{Improving the numerator sampling}
\label{sec:numsampling}

Apart from dealing with numerical instabilities, even for stable phase-space points there are at
least two possible applications of this tensorial reconstruction.

After the tensorial reconstruction is performed, we can feed the reconstructed numerator rather than the original one to the reduction algorithm. The added time needed to reconstruct the numerator can be compensated by a faster evaluation of the numerator during the reduction. For this reason, in the case of long expressions for the numerator function, processing the reconstructed tensor can be more efficient than the original numerator.

In order to address this issue, we performed a simple test: given a phase space point far from singularities, we consider a 6-point one-loop amplitude with a dummy numerator of rank $R$ made of one line of code by means of sums and powers of various scalar products.  
We can increase linearly the complexity of the numerator by adding identical copies of the same line and monitor the run time for the evaluation of the amplitude with or without a preliminary reconstruction of the tensor integrand $\langle N({q})\rangle$.
The results for the time ratio between the two methods for increasing sizes of the numerator are given in Table~\ref{rec:r46p}. 
It is interesting to observe that, in the case of rank $R=4$, we can easily gain a factor of 2 in the overall timing.

\begin{table}[ht]
\begin{center}
\begin{tabular}{|c|c|c|}
\hline 
\# Lines &  \multicolumn{2}{|c|}{Time ratio ``hybrid''/standard }\\
\hline 
N & Rank $=4$ & Rank $=6$\\ 
\hline
1      &    1.3&    1.6 \\
10     &    1.1&    1.4    \\
100    &    0.51&    0.85     \\
1000   &    0.30&     0.59   \\
10000  &    0.27&     0.55   \\
\hline
\end{tabular}
\caption{Timing for a rank 4 and rank 6 six-point function with a numerator of $N$ lines.
We display the ratio between the run times of the ``hybrid'' method and the 
standard reduction. 
In the ``hybrid'' method we compute the reconstructed tensor $\langle N({q})\rangle$ and use it for the reduction in place of the original numerator: already for a numerator of about $N=100$ lines, we get a significant improvement in computation time.} \label{rec:r46p}
\end{center}
\end{table}

As a further application, the method allows for a sampling based on real values of the integration momentum: this feature can be used to facilitate the adaptation of tree-level generators to the task of producing one-loop numerator functions.

\subsection{Implementation with \samurai{} and \golem{}}

For all the examples and calculations contained in this paper, the new techniques 
have been implemented using {\samurai} for a $d$-dimensional integrand-level reduction, and 
\golem{} for the evaluation of the tensor integrals.
However the validity of the methods presented goes beyond our specific implementation
and can be used in several alternative frameworks.

Coming back to our implementation, it required updates within \samurai{} 
which will be part of a future release of the package. In particular:

\noindent
{\bf Rescue System:} Unstable points, detected by means of one of the available tests described above, will be automatically
reprocessed using the tensorial decomposition. The evaluation of tensor integrals is performed by \golem{}.

\noindent
{\bf Tensorial numerator:} A flag will allow to reconstruct the tensorial integrand $\langle N({q})\rangle$, to be processed by the reduction algorithm in place of the original numerator. This is useful in the case of long expressions for the numerator function.

\section{Conclusions and Outlook}
\label{sec:end}

We presented a new approach to the reduction of one-loop amplitudes in which we
reconstruct the tensorial expression of the scattering 
amplitudes by means of a sampling in the integration momentum at the integrand level.

This approach could be used within existing techniques for the reduction of
one-loop multi-leg amplitudes in several ways; 
for example to deal with numerically unstable points,  
to allow for a sampling of the numerator function based on real values of the integration momentum, and to optimize the numerical reduction in the case of long expressions for the numerator functions.

We also illustrated how the existing reconstruction tests can be employed efficiently to detect potentially unstable phase space points. 
We believe that this method can represent a viable option as a
``rescue-system'' to deal with unstable phase space configurations:
the increase in computation time required by the evaluation of 
tensorial ``master" integrals appears to be less costly than 
reprocessing the unstable points with quadruple- or multi-precision routines. 

The tensorial reconstruction has been implemented and successfully tested within \samurai{} and \golem{}
and will be part of the next release of these programs.

\section*{Acknowledgments}

We would like to acknowledge the kind hospitality and support of the Theory Department at CERN
for the workshop ``\emph{Perturbative higher-order effects at work at the LHC - HO10}'', 
during which this project has been discussed and developed.
The authors are indebted to P. Mastrolia for collaboration at the beginning of this project, enlightening discussions
during its development and a critical reading of the manuscript.
F.T. gratefully acknowledges R. Pittau for discussions.
G.H. would like to thank the Pauli Center for Theoretical Studies, ETH/University of Z\"urich, for hospitality and support while part of this work has been completed.
G.H. is supported by the British Science and Technology Facilities Council (STFC).
The work of G.O. was supported by the NSF Grant PHY-0855489 and PSC-CUNY Award~60041-39~40; T.R. has been supported by the Foundation FOM, project FORM 07PR2556.

\appendix
\section{Combinatorics for tensor components}

The following table lists the block sizes and multiplicities
of the systems for $d=4$ up to rank~6. 
As explained in section \ref{sec:tensred}, for rank $r$, at each level $i$, 
we have $l_i^r=\left(4\atop i\right)\left(r\atop i\right)$ coefficients.
The numbers sum up to
$n(R)=\sum_{i=0}^R l_i^r=\sum_{r=0}^R n_r$.

\begin{subequations}
\begin{align}
n(0) =1&=1\times\left(4\atop0\right)\\
n(1) =5&=
	 1\times\left(4\atop0\right)
	+1\times\left(4\atop1\right)\\
n(2) =15&=
	 1\times\left(4\atop0\right)
	+2\times\left(4\atop1\right)
	+1\times\left(4\atop2\right)\\
n(3) =35&=
	 1\times\left(4\atop0\right)
	+3\times\left(4\atop1\right)
	+3\times\left(4\atop2\right)
	+1\times\left(4\atop3\right)\\
n(4) =70&=
	 1\times\left(4\atop0\right)
	+4\times\left(4\atop1\right)
	+6\times\left(4\atop2\right)
	+4\times\left(4\atop3\right)
	+1\times\left(4\atop4\right)\\
n(5) =126&=
	 1\times\left(4\atop0\right)
	+5\times\left(4\atop1\right)
	+10\times\left(4\atop2\right)
	+10\times\left(4\atop3\right)
	+5\times\left(4\atop4\right)\\
n(6) =210&=
	 1\times\left(4\atop0\right)
	+6\times\left(4\atop1\right)
	+15\times\left(4\atop2\right)
	+20\times\left(4\atop3\right)
	+15\times\left(4\atop4\right)
\end{align}
\end{subequations}
Hence, the largest system appears at rank six and is of size~20.

\bibliographystyle{utphys} 
\bibliography{MasterMORT.bib}

%
\end{document}